\documentclass[journal]{IEEEtran}
\usepackage{upgreek}
\usepackage{graphicx}
\usepackage{subfigure}
\usepackage{amsmath,bm}
\usepackage{amssymb}

\usepackage{pifont}
\usepackage{enumerate}

\usepackage{url}

\usepackage[square, comma, sort&compress, numbers]{natbib}

\usepackage{color}

\begin{document}
	
	\title{Duplex Self-Aligning Resonant Beam Communications and Power Transfer with Coupled Spatially Distributed Laser Resonator}
	\author{
		Mingliang Xiong,~\IEEEmembership{Member,~IEEE}, Qingwen Liu,~\IEEEmembership{Senior Member,~IEEE}, Hao Deng,~\IEEEmembership{Member,~IEEE}, \\ Gang Wang,~\IEEEmembership{Senior Member,~IEEE}, Gang Li,~\IEEEmembership{Member,~IEEE}, and Bin He,~\IEEEmembership{Senior Member,~IEEE}
		\thanks{
			
		}
		\thanks{This work was supported in part by the National Natural Science Foundation of China under Grant 62305019, 62371342, and U23B2059. (Corresponding Author: Qingwen Liu)} 
		\thanks{
			M. Xiong,  Q. Liu, and H. Deng
			are with the College of  Computer Science and Technology, Tongji University, Shanghai 201804, China (e-mail: mlx@tongji.edu.cn; qliu@tongji.edu.cn; denghao1984@tongji.edu.cn)
			
			G. Wang is with the State Key Lab of Intelligent Autonomous Systems, School of Automation, Beijing Institute of Technology, Beijing 100081, China (e-mail: gangwang@bit.edu.cn)
			
			G. Li and B. He are with  Shanghai Research Institute for Autonomous Intelligent Systems, Tongji University, Shanghai 201804, China. (lig@tongji.edu.cn; hebin@tongji.edu.cn)}
	}
	
	\maketitle
	
	\begin{abstract}
		Sustainable energy supply and high-speed communications are two significant needs for mobile electronic devices. This paper introduces a  self-aligning resonant beam system for simultaneous light information and power transfer~(SLIPT), employing a novel coupled spatially distributed resonator (CSDR). The system utilizes a resonant beam for efficient power delivery and a second-harmonic beam for concurrent data transmission, inherently minimizing echo interference and enabling bidirectional communication. Through comprehensive analyses, we investigate the CSDR's stable region, beam evolution, and power characteristics in relation to working distance and device parameters. Numerical simulations validate the CSDR-SLIPT system's feasibility by identifying a stable beam waist location for achieving accurate mode-match coupling between two spatially distributed resonant cavities and demonstrating its operational range and efficient power delivery across varying distances. The research reveals the system's benefits in terms of both safety and energy transmission efficiency. We also demonstrate the trade-off among the reflectivities of the cavity mirrors in the CSDR. These findings offer valuable design insights for resonant beam systems, advancing SLIPT  with significant potential for remote device connectivity.
	\end{abstract}
	
	
	\section{Introduction}\label{sec:intro}
	
	\IEEEPARstart{W}{ireless} power transfer (WPT)~\cite{WPT_review1, WPT_review2, WPT_review3}  can improve device portability by reducing cable dependence and potentially extending battery life, while optical wireless communication (OWC)~\cite{WOC_review1, WOC_review2} holds the potential to increase communication bandwidth. These advancements are becoming increasingly beneficial with the growth of technologies such as the Internet of Things (IoT)~\cite{IoT_devices}, augmented/virtual reality (AR/VR), and unmanned aerial vehicles (UAVs). 
	
	Simultaneous lightwave information and power transfer (SLIPT) synergizing WPT and OWC emerges as a significant approach to revolutionize wireless systems by enabling the concurrent and efficient transmission of both energy and data~\cite{SLIPT_concept1}. This integration is crucial for realizing truly interconnected devices, offering benefits such as improved system performance, cost reduction, simplified deployment, and enhanced user experiences. There are mainly two types of light source for SLIPT system: laser and light-emitting diode (LED). Laser-based SLIPT systems have distinct advantages due to their high directionality, extended range, and energy density, enabling efficient long-range power and data delivery \cite{laser_SLIPT_advantage}. However, existing laser-based SLIPT technologies face  challenges in beam alignment, tracking, and high-power safety \cite{a211117.01,laser_SLIPT_challenge1, laser_SLIPT_challenge2},i.e., high-power laser beams enhance energy delivery but also raise safety concerns regarding potential risks associated with free-space beam propagation and exposure.
	
	\begin{figure}[t]
		\centering
		\includegraphics[width=3.2in]{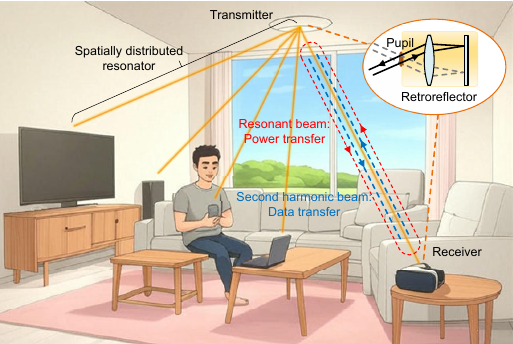}
		\caption{Demonstration of spatially distributed resonantor enabled simultaneous light information and power transfer}
		\label{fig:demo}
	\end{figure}
	
	To overcome the aforementioned challenges, a self-aligning SLIPT system based on resonant beam technologies was proposed~\cite{Xiong2022-kk}. This system is architected around a unique structure known as a spatially distributed resonator (SDR)~\cite{Liu2022-jc,Xiong2021-xs}.  As depicted in Fig.~\ref{fig:demo}, the SDR comprises two retroreflectors; one is integrated into the transmitter, while the other is incorporated within the receiver. Retroreflectors are characterized by their ability to reflect incident beams which pass through their pupil directly back towards their source. Taking this characteristic, laser oscillation is generated and sustained in the SDR's cavity, leveraging the optical gain provided by the laser crystal situated within the SDR. Recently, many works demonstrated the characteristics of SDR, which is also known as distributed cavity laser or alignment-free laser~\cite{Wang2021-oo,Zuo2023-fx,Zhang2022-bt,Lim2019-bs,Naqvi2023-og}. These studies provide a solid foundation for applications in WPT and OWC.
	
	In this paper, we propose a SLIPT system leveraging a coupled SDR~(CSDR) which comprises two tightly coupled spatially distributed resonant cavities~\cite{Sheng2023-rj,Deng2024-rc}.
	The primary intra-sub-resonator, located at the transmitter, is designed for optical oscillation and resonant beam generation.
	The secondary extra-sub-resonator establishes a self-aligning optical path connecting the transmitter and receiver, and specifically it performs like a Fabry-Perot~(F-P) cavity to provide a frequency-selective reflectivity for the primary resonator. The reflection peak of the secondary resonator effectively increases the reflectivity of the output mirror of the primary resonator.
	This configuration enhances the resonant beam power within the gain medium while simultaneously minimizing the power of the beam propagating in free space, which is crucial for improving both system safety and reducing propagation losses, especially under adverse atmospheric conditions.
	
	Besides, simultaneous duplex communication and power transfer are achieved through a dual-beam approach, utilizing the resonant beam for power delivery and a second harmonic beam, generated via second harmonic generation (SHG), for data transmission~\cite{Xiong2022-vh}. The wavelength separation between these beams inherently minimizes interference, while a time-division duplexing (TDD) scheme facilitates bidirectional data exchange without signal collision.
	
	The contributions of this work are as follows:
	\begin{itemize}
		\item A novel CSDR-based SLIPT system that achieves efficient energy transfer with inherent self-alignment and safety improvement.
		\item Implementation of second-harmonic beam for data communication, effectively decoupling data transmission from power transfer and mitigating interference.
		\item Development of an analytical model for the CSDR, illustrating the relationship between cavity stability and system parameters.
		\item Identification of a stable beam waist location within the CSDR, invariant to working distance, enabling robust mode-matching and large-range operation without continuous adjustments.
		\item Comprehensive system model and performance analysis providing a theoretical framework for the design and optimization of CSDR-based SLIPT systems.
	\end{itemize}
	
	\begin{figure*}[t]
		\centering
		\includegraphics[width=4.8in]{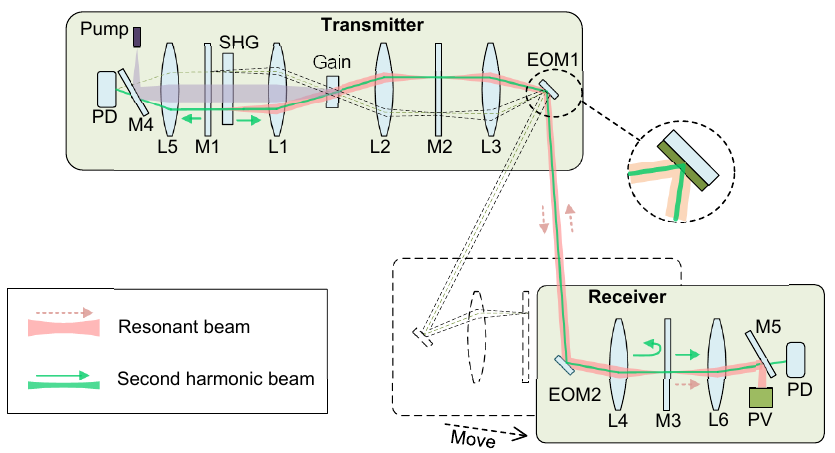}
		\caption{System design of the duplex simultaneous light information and power transfer system based on a coupled spatially distributed resonator}
		\label{fig:sysdesign}
	\end{figure*}
	
	The remainder of this paper is structured as follows: Section \ref{sec:design} details the system design, elaborating on the architecture of the duplex self-aligning resonant beam system. Section \ref{sec:model} presents the analytical system model, providing a theoretical foundation for performance analysis. Numerical results and discussions are presented in Section \ref{sec:result}, validating the proposed system and exploring its characteristics under various conditions. Finally, we conclude in Section \ref{sec:con}.
	
	\section{The Proposed System}
	\label{sec:design}

	As illustrated in Fig. \ref{fig:sysdesign}, the proposed  CSDR-SLIPT system consists of two main devices: the transmitter and the receiver, and they form a CSDR. The CSDR is coupled with two sub-resonators, i.e., an intra-sub-resonator for optical oscillation and amplification and an extra-sub resonator for self-alignment. The CSDR is designed to generate a self-aligning resonant beam for efficient power transfer. A second-harmonic beam is generated for data transfer through a SHG process, which can escape the speciafically-designed wavelength-selective cavity mirrors to avoid echo interference. If we modulate the resonant beam directly without SHG, the oscillation operation within the resonator leads to signal overlapping, i.e., the echo interference. Finally, a photovoltaic panel~(PV) is employed for energy harvesting and two photodiodes~(PDs) are used for information receiving.
	
	\subsection{Coupled Spatially Distributed Laser Resonator}
	The CSDR consists of mirrors (M1, M2, and M3), lenses (L1, L2, L3, and L4), a gain medium, and a SHG crystal. Different from a common laser resonator that comprises generally two mirrors, the main components here are retroreflectors, such as cat's eye and corner cube. A cat-eye  retroreflector is a mirror-lens pair, and in this system they are M1-L1, L2-M2, M2-L3, and L4-M3. The retroreflectors M1-L1 and L2-M2 form the primary resonant cavity, i.e., the intra-sub-resonator. The retroreflectors M2-L3 and L4-M3 form the secondary resonant cavity, i.e., the extra-sub-resonator. Two sub-resonators are coupled with each other through a partially reflective mirror M2. The extra-sub-resonator serves as a Fabry-Perot (F-P) cavity, exhibiting a frequency comb in its transmittance spectrum. This frequency-selective transmittance, combined with the mode competition principle, leads the resonant beam wavelength to be generated at the reflectivity peak. The gain medium, such as Nd:YVO$_4$ crystals, pumped by an external source, provides power amplification to maintain the laser oscillation. More details on SDR resonant beam systems can be found in previous studies~\cite{Liu2022-jc,Xiong2021-xs}.
	
	The CSDR is designed to be spatially distributed, meaning that the cavity of the extra-sub-resonantor is separated by a significant working distance between the transmitter and the receiver. This spatial distribution ability is enabled by the retroreflectors, allowing for a long resonant cavity length and the self-alignment capability. The coupled configuration is designed to reduce the propagation loss of the resonant beam in the air, which in turn improves the  charging power for the receiver's circuit and battery.
	
	\subsection{Beam Generation for SLIPT}
	The CSDR generates two laser beams simultaneously: a resonant beam and a second harmonic beam. The resonant beam is also referred to as the fundamental beam because it generates the second harmonic beam.  The resonant beam is used to transfer power and maiantain the resonance. To optimize the beam waist and divergence and ensure the cavity stability, resonator parameters, including mirror curvatures and lens focal lengths, are carefully chosen. The fundamental beam wavelength $\lambda_{\nu}$ is around $1064~{\rm nm}$, determined by the gain spectrum and the pump wavelength. The second harmonic beam, with a wavelength of $\lambda_{2\nu} = \lambda_{\nu}/2=532~{\rm nm}$, is generated through the SHG process. The SHG crystal is located close to mirror M3, which is phase-matched for efficient frequency doubling of the fundamental beam. 
	
	M3 is specifically designed to have different reflectivities at $\lambda_{\nu}$ and $\lambda_{2\nu}$. The reflectivity of M3 at $\lambda_{\nu}$ is denoted by $R_{\rm M3}$, while that at $\lambda_{2\nu}$ is denoted by $R'_{\rm M3}$. In CSDR, reflectivity $R_{\rm M3}$ is small (e.g., $10\%$) compared with a conventional resonator (usually $> 80\%$), so that almost $90\%$ power is released for receiver charging and only 10\% is reflected back to the transmitter to maintain the self-aligning resonance. The choice of reflectivity $R'_{M3}$  at $\lambda_{\rm 2\nu}$ depends on the power splitting strategy, which leads to a tradeoff between uplink and downlink communications. Specifically, a lower $R'_{\rm M3}$ allows more second harmonic power to be transmitted to the receiver for downlink communication, while a higher $R'_{\rm M3}$ reflects more power back to the transmitter for uplink communication.
	
	\subsection{Communications and Energy Harvesting}
	At the receiver, the fundamental beam and the second harmonic beam are separated by dichroic mirror M5 which is designed to reflect the second harmonic beam (information signal) towards photodetector PD2 and transmit the fundamental beam (power beam) towards the photovoltaic (PV) panel. The fundamental beam is focused by lens L6 onto a photovoltaic (PV) panel for energy harvesting. The PV panel converts the received optical power into current for battery charging.
	
	The second harmonic beam is modulated by an electro-optic modulator (EOM1) for downlink data transmission. For uplink data transmission, a portion of the second-harmonic beam is reflected back to the transmitter by M3 through the beam path of downlink communication, and this reflected beam is modulated by another electro-optic modulator (EOM2) at the receiver. The uplink second harmonic beam is detected by another photodetector PD1 equipped in the transmitter. Since the downlink and the uplink share the same path, the modulation of EOM1 and EOM2 should occur at different times, thereby a time-division duplexing (TDD) scheme should be adopted.

	\section{System Model}
	\label{sec:model}
	
\begin{figure*}[t]
		\centering
		\includegraphics[width=4in]{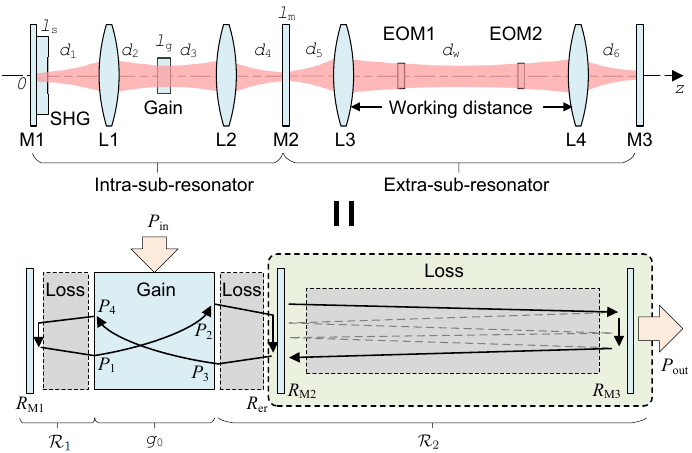}
		\caption{System model for analysis of the coupled spatially distributed resonator}
		\label{fig:sysmodel}
	\end{figure*}

	To analyze the performance of the proposed CSDR-SLIPT system, we develop a system model that considers the beam propagation characteristics, laser resonator dynamics, second harmonic generation, wireless power transfer, and optical communication links.  Figure \ref{fig:sysmodel} illustrates the system model, highlighting the structures, components, and key parameters.
	
	In our system model, we make several assumptions to simplify the analysis while capturing the essential system behaviors. We assume that the reflectivities, $R_{\rm ar}$, and transmissivities, $T_{\rm ar}$, of anti-reflecting (AR) coatings are device-independent, implying consistent values across devices. The above assumption also applies to those of high-reflecting (HR) coatings, i.e., $R_{\rm hr}$ and $T_{\rm hr}$. We also assume that the reflectivities and transmissivities are wavelength-independent (e.g., L1~--~L6), unless otherwise specified (e.g., M1~--~M5, EOM1, and EOM2). These approximations are reasonable for high-quality optical coatings and components operating within a narrow wavelength band. However, in practice, coating performance may vary slightly between devices and wavelengths, which could introduce minor deviations from the model.
	
	\subsection{Fundamental Beam Generation}
	
	To analyze the resonator stability and the intra-cavity resonant beam radius, we employ the ABCD matrix analysis~\cite{Magni1987-oz}.  The single-pass ray-transfer matrix of the coupled resonator, denoted as $\left[A_{\rm r}, B_{\rm r}; C_{\rm r}, D_{\rm r}\right]$, represents the transformation of the beam parameters after propagating from M1 to M3 in the CSDR. This matrix is calculated by multiplying the ray-transfer matrices of each optical component in the cavity in a reverse sequence to the propagation direction:
	
	\begin{equation}
		\begin{aligned}
			\begin{bmatrix}
				A_{\rm r}&B_{\rm r}\\ C_{\rm r}&D_{\rm r}
			\end{bmatrix}
			=&\begin{bmatrix}
				1&d_6 \\ 0&1
			\end{bmatrix}
			\begin{bmatrix}
				1&0 \\ -\frac{1}{f_{\rm L4}}&1
			\end{bmatrix}
			\begin{bmatrix}
				1&D_{\rm r} \\ 0&1
			\end{bmatrix}
			\begin{bmatrix}
				1&0 \\ -\frac{1}{f_{\rm L3}}&1
			\end{bmatrix}
			\\
			&\begin{bmatrix}
				1&d_5 \\ 0&1
			\end{bmatrix}
			\begin{bmatrix}
				1&\frac{l_{\rm m}}{n_{\rm m}} \\ 0&1
			\end{bmatrix}
			\begin{bmatrix}
				1&d_4 \\ 0&1
			\end{bmatrix}
			\begin{bmatrix}
				1&0 \\ -\frac{1}{f_{\rm L2}}&1
			\end{bmatrix}
			\\
			&\begin{bmatrix}
				1&d_3 \\ 0&1
			\end{bmatrix}
			\begin{bmatrix}
				1&\frac{l_{\rm g}}{n_{\rm g}} \\ 0&1
			\end{bmatrix}
			\begin{bmatrix}
				1&d_2 \\ 0&1
			\end{bmatrix}
			\begin{bmatrix}
				1&0 \\ -\frac{1}{f_{\rm L1}}&1
			\end{bmatrix}
			\\
			&\begin{bmatrix}
				1&d_1 \\ 0&1
			\end{bmatrix}
			\begin{bmatrix}
				1&\frac{l_{\rm s}}{n_{\rm s}} \\ 0&1
			\end{bmatrix},
		\end{aligned}
	\end{equation}
	where $d_1$, $d_2$, $d_3$, $d_4$, $d_5$, and $d_6$ represent the distances between optical components as shown in Fig. \ref{fig:sysmodel}. $d_{\rm w}$ is the wireless transmission distance (working distance). $f_{\rm L1}$, $f_{\rm L2}$, $f_{\rm L3}$, and $f_{\rm L4}$ are the focal lengths of lenses L1, L2, L3, and L4, respectively. $l_{\rm s}$ and $n_{\rm s}$ are the thickness and refractive index of the SHG crystal, respectively. $l_{\rm g}$ and $n_{\rm g}$ are the thickness and refractive index of the gain medium, respectively. $l_{\rm m}$ and $n_{\rm m}$ are the thickness and refractive index of the partially reflective mirror M2, respectively. 
	
	To analyze the resonator stability, we define the stability parameters $g_1^*$ and $g_2^*$ and the equivalent cavity length parameter $L^*$ based on the elements of the single-pass ray-transfer matrix:
	\begin{equation}
		\begin{aligned}
			g_1^*&=A_{\rm r},\\
			g_2^*&=D_{\rm r},\\
			L^*&=B_{\rm r}.
		\end{aligned}
	\end{equation}
	A stable resonator, which supports the formation of a Gaussian beam mode, should satisfy the stability criterion~\cite{a181221.01}:
	\begin{equation}
		0<g_1^*g_2^*<1 ~~\text{or} ~~g_1^*=g_2^*=0.
		\label{equ:stab}
	\end{equation}
	This stable region ensures that the resonant beam remains bounded in the resonant cavity and does not diverge after multiple round trips.  Resonators operating outside this stable region will exhibit high losses and prevent efficient laser oscillation. The condition $g_1^*=g_2^*=0$ represents a special case known as the confocal configuration, which is also a stable state.
	
	The resonant beam may contains many transverse modes. The radius of the fundamental mode, i.e., the TEM$_{00}$ mode, on the leftmost resonator mirror, M1, denoted as $w_{00}(0)$, can be calculated using the resonator parameters and the laser wavelength $\lambda_{\nu}$~\cite{a181221.01,Herziger1984-kd}:
	\begin{equation}
		w_{00}(0)=\sqrt{\frac{\lambda_{\nu}|L^*|}{\pi}\sqrt{\frac{g_2^*}{g_1^*(1-g_1^*g_2^*)}}}.
	\end{equation}
	The mode radius $w_{00}(0)$ at mirror M1 is a key parameter that can be used to determine the beam size and divergence throughout the resonator cavity, and it is essential for calculating the output beam power of the resonator.
	
	Next, we deduce the q-parameters of the resonator. The q-parameter is a complex parameter that encapsulates both the beam radius and wavefront curvature of a Gaussian beam~\cite{a181221.01}. It is a convenient tool for tracking beam radius evolution as it propagates through optical systems. The q-parameter at mirror M1, $q(0)$ is given by:
	\begin{equation}
		q(0)=\frac{j\pi w_{00}^2(0)}{\lambda_{\nu}},
	\end{equation}
	where $j$ is the imaginary unit.
	
	According to the ABCD law of Gaussian beams, we can deduce the q-parameters at any location $z$ along the beam propagation path using the ray-transfer matrix from mirror M1 ($z=0$) to the location $z$. The q-parameter at location $z$, $q(z)$, is transformed according to~\cite{a181221.01}:
	\begin{equation}
		q(z)=\frac{A_{\rm t}(z)q(0)+B_{\rm t}(z)}{C_{\rm t}(z)q(0)+D_{\rm t}(z)},
	\end{equation}
	where $A_{\rm t}(z)$, $B_{\rm t}(z)$, $C_{\rm t}(z)$, and $D_{\rm t}(z)$ are the elements of the ray-transfer matrix $\mathbf{M}_{\rm t}(z)$
	from M1 ($z=0$) to the location $z$. This matrix, $\mathbf{M}_{\rm t}(z)$, can be formulated as a piecewise function based on the location $z$ along the optical path:
	\begin{equation}
		\begin{aligned}
			&\mathbf{M}_{\rm t}(z)=\begin{bmatrix}
				A_{\rm t}(z)&B_{\rm t}(z) \\ C_{\rm t}(z)&D_{\rm t}(z)
			\end{bmatrix} =
			\\ &
			\begin{cases}
				&\mathbf{m}_{\rm n}(z,n_{\rm s}),~~z\in[0,z_{\rm sr}]
				\\ &
				\mathbf{m}_{\rm n}(z-z_{\rm sr},1)
				\mathbf{M}_{\rm t}(z_{\rm sr})
				,~~z\in(z_{\rm sr},z_{\rm L1}]
				\\ &
				\mathbf{m}_{\rm n}(z-z_{\rm L1},1)
				\mathbf{m}_{\rm f}(f_{\rm L1})
				\mathbf{M}_{\rm t}(z_{\rm L1})
				,~~z\in(z_{\rm L1},z_{\rm gl}]
				\\ &
				\mathbf{m}_{\rm n}(z-z_{\rm gl},n_{\rm g})
				\mathbf{M}_{\rm t}(z_{\rm gl})
				,~~z\in(z_{\rm gl},z_{\rm gr}]
				\\ &
				\mathbf{m}_{\rm n}(z-z_{\rm gr},1)
				\mathbf{M}_{\rm t}(z_{\rm gr})
				,~~z\in(z_{\rm gr},z_{\rm L2}]
				\\ &
				\mathbf{m}_{\rm n}(z-z_{\rm L2},1)
				\mathbf{m}_{\rm f}(f_{\rm L2})
				\mathbf{M}_{\rm t}(z_{\rm L2})
				,~~z\in(z_{\rm L2},z_{\rm ml}]
				\\ &
				\mathbf{m}_{\rm n}(z-z_{\rm ml},n_{\rm m})
				\mathbf{M}_{\rm t}(z_{\rm ml})
				,~~z\in(z_{\rm ml},z_{\rm mr}]
				\\ &
				\mathbf{m}_{\rm n}(z-z_{\rm mr},1)
				\mathbf{M}_{\rm t}(z_{\rm mr})
				,~~z\in(z_{\rm mr},z_{\rm L3}]
				\\ &
				\mathbf{m}_{\rm n}(z-z_{\rm L3},1)
				\mathbf{m}_{\rm f}(f_{\rm L3})
				\mathbf{M}_{\rm t}(z_{\rm L3})
				,~~z\in(z_{\rm L3},z_{\rm L4}]
				\\ &
				\mathbf{m}_{\rm n}(z-z_{L4},1)
				\mathbf{m}_{\rm f}(f_{\rm L4})
				\mathbf{M}_{\rm t}(z_{\rm L4})
				,~~z\in(z_{\rm L4},z_{\rm M3}]
			\end{cases}
		\end{aligned}
		\label{equ:piswf}
	\end{equation}
	where $z_{\rm sr}=l_{\rm s}$, $z_{\rm L1}=z_{\rm sr}+d_1$, $z_{\rm gl}=z_{\rm L1}+d_2$, $z_{\rm gr}=z_{\rm gl}+l_{\rm g}$, $z_{\rm L2}=z_{\rm gr}+d_3$, $z_{\rm ml}=z_{\rm L2}+d_4$, $z_{\rm mr}=z_{\rm ml}+l_{\rm m}$, $z_{\rm L3}=z_{\rm mr}+d_5$, $z_{\rm L4}=z_{\rm L3}+d_{\rm w}$, and $z_{\rm M3}=z_{\rm L4}+d_6$.
	The matrices $\mathbf{m}_{\rm n}(z,n)$ and $\mathbf{m}_{\rm f}(f)$ represent the ray-transfer matrices for propagation through a refractive medium, and refraction through a thin lens, respectively; they are given by:
	\begin{equation}
		\mathbf{m}_{\rm n}(z,n)=
		\begin{bmatrix}
			1&\frac{z}{n}\\0&1
		\end{bmatrix},
		\quad\mathbf{m}_{\rm f}(f)=
		\begin{bmatrix}
			1&0\\-\frac{1}{f}&1
		\end{bmatrix},
	\end{equation}
	where $z$ is the propagation distance, $n$ is the refractive index of the medium (here we assume refractive index of free space is 1), and $f$ is the focal length of the lens. The piecewise function in \eqref{equ:piswf} is formulated to model the overall matrix of a optical system from 0 to $z$, which is derived from multiplying different component matrices across various transmission distances.
	
	The TEM$_{00}$ mode radius of the fundamental beam at any position $z$, $w_{00}(z)$, can be obtained from the q-parameter $q(z)$ using the following formula~\cite{a181221.01}:
	\begin{equation}
		w_{00}(z)=\sqrt{-\frac{\lambda_{\nu}}{\pi \Im[1/q(z)]}},
		\label{equ:w00}
	\end{equation}
	where $\Im[\cdot]$ denotes the imaginary part of a complex number. Equation \eqref{equ:w00} allows us to calculate the resonant beam radius at any position $z$ within the resonator, based on the q-parameter $q(z)$ at that location; that is
	\begin{equation}
		w(z)= Mw_{00}(z),
		\label{equ:M2}
	\end{equation}
	where $M=a_{\rm g}/w_{00}(z_{\rm gl})$ is the beam propagation factor defined by the gain medium radius $a_{\rm g}$ and the TEM$_{00}$ mode radius at the gain medium, assuming the gain medium aperture is the smallest aperture in the resonant beam path, and therefore, dominates the diffraction loss. The beam propagation factor $M$ accounts for the beam expansion or contraction as it propagates through the resonator, relative to the beam size at the gain medium.
	
	To evaluate the resonant beam power, we should convert the physical cavity configuration into a canonical laser power cycling model which contains only two loss regions and a gain region, as shown in Fig.~\ref{fig:sysmodel}. Then, based on the Rigrod's analysis method \cite{a190318.02}, the output power of the CSDR, $P_{\rm out}$, can be expressed as~\cite{Xiong2021-xs,a181224.01}:
	\begin{equation}
		P_{\rm out}=T_{\rm 2o}P_2=\eta_{\rm slop}[P_{\rm in}-P_{\rm th}],
		\label{equ:Pout}
	\end{equation}
	where $\eta_{\rm slop}$ represents the slope efficiency, and $P_{\rm th}$ is the threshold pump power. Note that, here $P_{\rm out}=0$ when $P_{\rm in}<P_{\rm th}$. These parameters are given by:
	\begin{equation}
		\left\{
		\begin{aligned}
			\eta_{\rm slop}&=\frac{ w_{\rm g}^2 \eta_{\rm c}T_{\rm 2o}}{a_{\rm g}^2\left( 1+\sqrt{\frac{\mathcal{R}_2}{\mathcal{R}_1}}\right)\left(1-\sqrt{\mathcal{R}_1\mathcal{R}_2}\right) },
			\\
			P_{\rm th}&=\frac{\pi a_{\rm g}^2 I_{\rm s}}{\eta_{\rm c}}\ln\frac{1}{\sqrt{\mathcal{R}_1\mathcal{R}_2}},
		\end{aligned}
		\right.
	\end{equation}
	where $w_{\rm g}$ is the radius of the resonant beam in the gain medium, and $a_{\rm g}$ is the radius of the pump beam which is assumed equal to the gain aperture. We approximate $w_{\rm g} = a_{\rm g}$ due to the physical reality that the effective beam radius for amplification is constrained by the gain medium size. If the TEM$_{00}$ mode is narrower, higher-order modes appears, expanding the beam spot. Other modes whose radius is greater than the gain medium aperture can not been generated as they will be truncates by the aperture. $\eta_{\rm c}$ is the combined efficiency for pump power propagation and absorption by the gain medium. $I_{\rm s}$ is the saturation intensity of the gain medium. $T_{\rm 2o}=P_{\rm out}/P_2$ represents the combined transmittance from the right surface of the gain medium to the right surface of the output mirror M3 (from $P_2$ to $P_{\rm out}$), accounting for the losses in lens L2 and the extra-sub-resonator:
	\begin{equation}
		T_{\rm 2o}=T_{\rm gr}T_{\rm L2}T_{\rm er},
	\end{equation}
	where $T_{\rm er}$ is the effective transmissivity of the extra-sub-resonator which operates as a F-P cavity.
	$\mathcal{R}_1$ and $\mathcal{R}_2$ are the equivalent reflectivities of the left and right ends of the gain medium, respectively, considering the cascaded optical components; that is:
	\begin{equation}
		\left\{
		\begin{aligned}
			\mathcal{R}_1&=T_{\rm diff}T_{\rm gl}^2T_{\rm L1}^2T_{\rm s}^2 R_{\rm M1},
			\\
			\mathcal{R}_2&=T_{\rm diff}T_{\rm gr}^2T_{\rm L2}^2 R_{\rm er},
		\end{aligned}
		\right.
	\end{equation}
	where $R_{\rm M1}$ is the reflectivity of mirror M1. $T_{\rm diff}$ is the diffraction loss factor which contributes an attenuation to the resonant beam power as it propagation in the cavity. $R_{\rm er}$ is the equivalent reflectivity of the extra-sub-resonator (formed by M2, M3, L3, L4, and the air gap $d_{\rm w}$) in the resonant state. The effective transmissivity of the SHG crystal at the fundamental frequency is expressed as:
	\begin{equation}
		T_{\rm s}=(1-\eta_{\rm shg})T_{\rm sl} T_{\rm sr},
	\end{equation}
	where $\eta_{\rm shg}$ is the SHG efficiency, and $T_{\rm sl}$ and $T_{\rm sr}$ are the transmissivities of the left and right surfaces of the SHG crystal, respectively. Similarly, $T_{\rm gl}$ and $T_{\rm gr}$ are the transmissivities of the end surfaces of the gain medium. $T_{\rm L1}$ and $T_{\rm L2}$ are the transmissivities of lenses L1 and L2, respectively. Note that: $\eta_{\rm shg}$ is related to the fundamental beam power for SHG, $P_{\nu}$ (detailed in the next subsection), so we should at first come out $\eta_{\rm shg}$ by \eqref{equ:etashg}.
	
	Diffraction loss within the resonator arises at each aperture defined by the components, including lenses, mirrors, and crystals, which is intrinsically linked to the wave nature of light. Nevertheless, the diffraction loss is predominantly governed by the minimum aperture among all. In our resonator, the aperture of the gain medium is much smaller than other components. Therefore, the diffraction loss factor (single-pass) primarily accounts for the gain medium aperture, and it is approximated by:
	\begin{equation}
		T_{\rm diff}=1-\exp\left[-2\left(\frac{a_{\rm g}}{w_{00}(z_{\rm gl})}\right)^2\right].
		\label{equ:Tdiff}
	\end{equation}
	It should be noted that this formula is of limited accuracy, as it only considers the fundamental Gaussian mode and assumes the gain medium to be a thin aperture. For a more accurate solution, we should use a numerical method, such as  Fox-Li method~\cite{a201125.01}. Nevertheless, this approximation is sufficient to be use \eqref{equ:Tdiff} in evaluating the output power and the maximum transmission distance because the diffraction loss remains close to zero when varying over a wide working range and deteriorates rapidly as the resonant cavity approaches the margin of the stable region.
	
	The equivalent transmissivity of the extra-sub-resonator, $T_{\rm er}$, is deduced as~\cite{Ismail2016-ix}:
	\begin{equation}
		T_{\rm er}(\varphi)=\frac{(1-R_{\rm M2})(1-R_{\rm M3})T_{\rm ier}}{1+R_{\rm M2}R_{\rm M3}T_{\rm ier}^2+2T_{\rm ier}\sqrt{R_{\rm M2}R_{\rm M3}}\cos(\varphi)},
		\label{equ:Ter}
	\end{equation}
	where $R_{\rm M2}$ and $R_{\rm M3}$ are the reflectivities of mirrors M2 and M3, respectively. $T_{\rm ier}=T_{\rm air}T_{\rm L3}T_{\rm L4}R_{\rm e1}R_{\rm e2}$ represents the internal single-pass  transmittance through the optical lenses and air gap in the extra-sub-resonator, where $T_{\rm air}=e^{-\alpha_{\rm air}d_{\rm w}}$ ($\alpha_{\rm air}$ is the attenuation coefficient of air), and $T_{\rm L3}$ and $T_{\rm L4}$ are the transmissivities of lenses L3 and L4, respectively. $\varphi = 4\pi (d5+d_{\rm w}+d6)/\lambda_{\nu}$ is the phase factor related to the air gap length. In the formula of $T_{\rm ier}$, $R_{\rm e1}$ and $R_{\rm e2}$ represent the reflectivities of EOM1 and EOM2, respectively, assuming that the modulators also introduce some reflection loss to the resonant beam.
	
	Similarly, the equivalent reflectivity of the extra-sub-resonator, $R_{\rm er}$, which is essentially a F-P cavity with internal loss, is obtained as:
	\begin{equation}
		R_{\rm er}(\varphi)=\frac{R_{\rm M2}+R_{\rm M3}T_{\rm ier}^2+2T_{\rm ier}\sqrt{R_{\rm M2}R_{\rm M3}}\cos(\varphi)}{1+R_{\rm M2}R_{\rm M3}T_{\rm ier}^2+2T_{\rm ier}\sqrt{R_{\rm M2}R_{\rm M3}}\cos(\varphi)},
		\label{equ:Rer}
	\end{equation}
	
	\begin{figure}[t]
		\centering
		\includegraphics[width=3.2in]{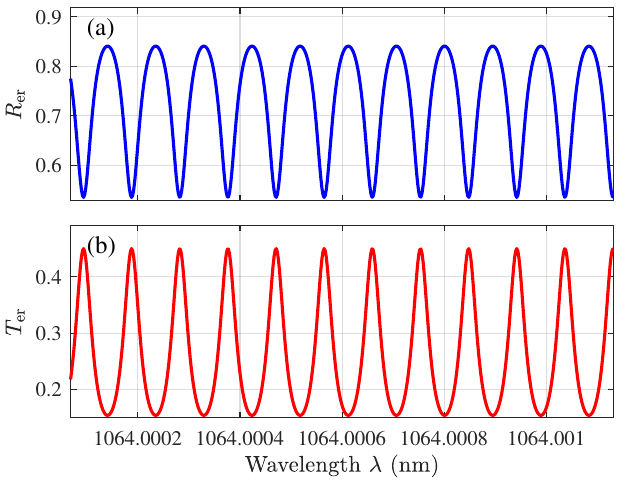}
		\caption{Reflectance (a) and  transmittance (b) spectra of the extra-sub-resonator, measured for an incident beam at M2 within the intra-sub-resonator.}
		\label{fig:fpspecrtum}
	\end{figure}
	
	Figure~\ref{fig:fpspecrtum} shows the reflectance ($R_{\rm er}$) spectrum and the transmittance ($T_{\rm er}$) spectrum of the extra-sub-resonator, which exhibits a comb-like structure with multiple peaks and valleys. In a CSDR, the resonant frequency corresponds to the minimum cavity loss, observed at the peaks of $R_{\rm er}(\varphi)$. This resonance condition is satisfied when $\cos(\varphi)=1$. Under this resonant condition, equations \eqref{equ:Ter} and \eqref{equ:Rer} are simplified to:
	\begin{equation}
		\begin{aligned}
			\hat{T}_{\rm er}&=\frac{(1-R_{\rm M2})(1-R_{\rm M3})T_{\rm ier}}{(1 + T_{\rm ier}\sqrt{R_{\rm M2}R_{\rm M3}})^2},\\
			\hat{R}_{\rm er}&=\frac{(\sqrt{R_{\rm M2}}+T_{\rm ier}\sqrt{R_{\rm M3}})^2}{(1+T_{\rm ier}\sqrt{R_{\rm M2}R_{\rm M3}})^2}.
		\end{aligned}
	\end{equation}
	
	\subsection{Second Harmonic Generation}
	The second-harmonic beam power can be deduced from the leftward-propagating fundamental beam power ($P_4$ in Fig.~\ref{fig:sysmodel}) that leaves the gain medium. Because of the symmetry of the equivalent power cycling model, as shown in Fig.~\ref{fig:sysmodel}, $P_4$ can be derived by reversing the positions of $\mathcal{R}_1$ and $\mathcal{R}_2$ in the formula of $P_2$. Similar to the output power formula~\eqref{equ:Pout}, the fundamental beam power incident on the SHG crystal, $P_{\rm \nu}$, can be expressed as:
	\begin{equation}
		\left\{
		\begin{aligned}
			P_{\rm \nu}&=\eta_{\rm slop}'[P_{\rm in}-P_{\rm th}],
			\\
			\eta_{\rm slop}'&=\frac{ w_{\rm g}^2 \eta_{\rm c}T_{\rm 2s}}{a_{\rm g}^2\left( 1+\sqrt{\frac{\mathcal{R}_1}{\mathcal{R}_2}}\right)\left(1-\sqrt{\mathcal{R}_2\mathcal{R}_1}\right) },
			\\
			P_{\rm th}&=\frac{\pi a_{\rm g}^2 I_{\rm s}}{\eta_{\rm c}}\ln\frac{1}{\sqrt{\mathcal{R}_1\mathcal{R}_2}},
		\end{aligned}
		\right.
	\end{equation}
	where $\eta_{\rm slop}'$ is the slope efficiency. $T_{\rm 2s}$ represents the equivalent tranmittance from the left surface of the gain medium to the SHG crystal, which is expressed as:
	\begin{equation}
		T_{\rm 2s}=T_{\rm gl}T_{\rm L1}.
	\end{equation}
	The single-pass SHG efficiency, $\eta_{\rm shg}$, in the SHG crystal can be approximated by \cite{a181218.01, Simon1997-fw}:
	\begin{equation}
		\eta_{\rm shg}=\frac{8\pi^2d_{\rm eff}^2 l_{\rm s}^2}{\varepsilon_0 c \lambda_{\nu}^2 n_{\rm s}^3}\cdot\frac{P_{\rm \nu}}{\pi w_{\rm s}^2},
		\label{equ:etashg}
	\end{equation}
	where $d_{\rm eff}$ and $l_{\rm s}$ is the effective nonlinear coefficient and length of the SHG crystal, respectively; $\varepsilon_0$ is the vacuum permittivity; $c$ is the light speed in vacuum; $n_{\rm s}$ is the refractive index of the SHG crystal at $\lambda_{\nu}$; and $w_{\rm s}$ is the radius of the fundamental beam in the SHG crystal. Here, we set the SHG crystal attached to mirror M1, so we have $w_{\rm s}=w(0)$.  This formula is derived under the assumption of plane wave approximation and small signal conversion efficiency. For more rigorous analysis, especially at high conversion efficiency, more complex models considering pump depletion and phase mismatch may be required. 
	
	In the CSDR, due to the bidirectional propagation of the fundamental beam, the SHG process generates two second-harmonic beams propagating in opposite directions. We denote their powers as $P_{\rm 2\nu}^-$ (leftward propagation towards M1) and $P_{\rm 2\nu}^+$(rightward propagation towards M3). These second harmonic powers can be estimated as:
	\begin{equation}
		\begin{aligned}
			P_{2\nu}^-&= \eta_{\rm shg}T_{\rm s}P_{\nu},\\
			P_{2\nu}^+&=\eta_{\rm shg}(1-\eta_{\rm shg})R_{\rm M1}T_{\rm s}^2P_{\nu},
		\end{aligned}
	\end{equation}
	where $T_{\rm s}$ denotes the  transmissivity of the SHG crystal surfaces which have AR coatings. The factor $(1-\eta_{\rm shg})$ in the expression for $P_{\rm 2\nu}^+$ accounts for the attenuation of the fundamental beam power $P_{\rm \nu}$ in the first pass through the SHG crystal, as the fundamental beam power in the positive direction (towards M3) originates from the leftward-travelling fundamental beam which has already experienced attenuation from SHG and M1. The leftward-travelling second harmonic beam, with power $P_{2\nu}^-$, remains unused, contributing to the uplink shot noise at PD1, which will be discussed in Section~\ref{sec:con}.
	
	\subsection{Wireless Power Transfer and Energy Harvesting}
	The optical power received by the PV panel, $P_{\rm pvh}$, is derived from the output power released by mirror M3, $P_{\rm out}$, considering the attenuation of the focusing lens L6 and the dichroic mirror M5, i.e., $P_{\rm pvh} = T_{\rm L6}R_{\rm M5}P_{\rm out}$.
	In photovoltaic conversion, the responsivity, $\rho_{\rm pv}$, quantifies the relationship between the incident received power and the generated photocurrent $I_{\rm pv}$. Hence, we have:
	\begin{equation}
		I_{\rm pv}=\rho_{\rm pv} T_{\rm pv}T_{\rm L6}R_{\rm M5} P_{\rm out},
	\end{equation}
	where $\rho_{\rm pv}$ (in A/W) depends on the PV panel material, structure, and the wavelength of the incident light. $T_{\rm pv}$ is the receiving efficiency of the PV panel, accounting for factors such as spectral mismatch and reflection losses.
	
	The relationship between the output voltage $V_{\rm d}$, output current $I_{\rm chg}$, and photo-generated current $I_{\rm pv}$ of the PV panel is described by the diode equation, considering the effects of series resistance $R_{\rm s}$ and shunt resistance $R_{\rm sh}$~\cite{a190923.01,Sera2007-jn}:
	\begin{equation}
		\left\{
		\begin{aligned}
			&I_{\rm chg}=I_{\rm pv}-I_0\left(e^{\frac{V_{\rm d}}{N_{\rm s}n_{\rm d}V_{\rm t}}}-1\right)-\frac{V_{\rm d}}{R_{\rm sh}},\\
			&V_{\rm d}= I_{\rm chg}(R_{\rm PL}+R_{\rm s}),
		\end{aligned}
		\right.
	\end{equation}
	where $I_0$ is the diode saturation current, $N_{\rm s}$ is the number of series-connected PV cells, $n_{\rm d}$ is the diode ideality factor, $V_{\rm t}=kT_{\rm t}/q_{\rm e}$ is the thermal voltage of the PV panel ($k$ is Boltzmann constant, $T_{\rm t}$ is temperature, $q_{\rm e}$ is elementary charge), and $R_{\rm PL}$ is the load resistance determined by the circuit. The diode equation models the non-linear current-voltage characteristics of the PV panel. The series resistance $R_{\rm s}$ represents internal losses of the PV panel, while the shunt resistance $R_{\rm sh}$ accounts for leakage current. 
	
	Typically, a PV charging circuit employs a maximum power point tracking (MPPT) algorithm to dynamically adjust the operating point of the PV panel and maximize the extracted electrical power. The MPPT process can be modeled as:
	\begin{equation}
		\begin{aligned}
			P_{\rm chg}=\max_{V_{\rm chg}}I_{\rm chg}V_{\rm chg},\\
			\mbox{s.t.} \left\{
			\begin{aligned}
				&V_{\rm chg}=R_{\rm PL}I_{\rm chg}\\
				&0\leq V_{\rm chg}\leq V_{\rm oc}
			\end{aligned}
			\right.
		\end{aligned}
	\end{equation}
	where $P_{\rm chg}$ is the charging power, $V_{\rm chg}$ is the charging voltage, $R_{\rm PL}$ is the load resistance, and $V_{\rm oc}$ is the open-circuit voltage of the PV panel. The constraints ensure that the charging voltage is matched to the load and remains within the operational limits of the PV panel.  MPPT algorithms, such as perturb and observe or incremental conductance, are commonly used to track the maximum power point by adjusting the load resistance $R_{\rm PL}$ to maximize the cahrging power $P_{\rm chg}$ under varying illumination conditions.
	
	\subsection{Down-Link and Up-link Data Transfer}
	
	In the down-link transmission, the received signal beam power at the photodetector PD2 in the receiver, $P_{\rm down}$, is derived from the rightward-propagating second-harmonic beam $P_{\rm 2\nu}^+$, considering the  transmission losses along the optical path:
	\begin{equation}
		\begin{aligned}
			P_{\rm down}=&T_{\rm pd2}T'_{\rm M5}T_{\rm L6}T'_{\rm M3}T_{\rm L4}R'_{\rm e2}T_{\rm air}R'_{\rm e1}T_{\rm L3}\\
			&\times T'_{\rm M2}T_{\rm L2}T_{\rm g}T_{\rm L1}P_{\rm 2\nu}^+
		\end{aligned}
	\end{equation}
	where $T_{\rm pd2}$ is the receiving efficiency of the photodetector PD2, $T'_{\rm M5}$, $T'_{\rm M3}$, and $T'_{\rm M2}$ are the transmissivities of mirrors M5, M3, and M2 at the second harmonic wavelength $\lambda_{2\nu}$, respectively. $T_{\rm L6}$, $T_{\rm L4}$, $T_{\rm L3}$, $T_{\rm L2}$, and $T_{\rm L1}$ are the transmissivities of lenses L6, L4, L3, L2, and L1, respectively. $T_{\rm g}$ is the transmissivity of the gain medium. $R'_{\rm e1}$ and $R'_{\rm e2}$ are the reflectivities of EOM1 and EOM2 at $\lambda_{\rm 2\nu}$. This equation accounts for the power loss at each optical component along the downlink path, from the second harmonic beam generation at the SHG crystal to the detection at PD2.
	
	In the up-link transmission, the received signal beam power at the photodetector PD1 in the transmitter, $P_{\rm up}$, is derived from the leftward-propagating second-harmonic beam $P_{\rm 2\nu}^-$ and the residual rightward-propagating second-harmonic beam that completes a round trip back to the transmitter:
	\begin{equation}
		\begin{split}
			P_{\rm up} &= C_1 C_2 P_{\rm 2\nu}^{+}, \\
			C_1 &= R'_{\rm M3}T_{\rm L4}R'_{\rm e2}T_{\rm air}R'_{\rm e1}T_{\rm L3} T'_{\rm M2}T_{\rm L2}T_{\rm g}T_{\rm L1}, \\
			C_2 &= T_{\rm pd1}T'_{\rm M4}T_{\rm L5}T'_{\rm M1}T_{\rm s} T_{\rm L1}T_{\rm g}T_{\rm L2}T'_{\rm M2}T_{\rm L3}T_{\rm L4}R'_{\rm e1}T_{\rm air}R'_{\rm e2}, 
		\end{split}
		\label{equ:Pup}
	\end{equation}
	where $C_1$ represents the combined tranmittance from M1 to M3, $C_2$ represents the combined transmittance for the partially back-reflected second-harmonic beam to PD1.  $T'_{\rm M4}$ and $T'_{\rm M1}$ are the transmissivities of mirrors M4 and M1 at the $\lambda_{2\nu}$, respectively. $T'_{\rm s}$ is the transmissivity of the SHG crystal at the second harmonic wavelength.  Other parameters are defined similarly to the downlink case. From the equation~\eqref{equ:Pup}, we can see that $R'_{\rm M3}$ serves as the power allocator between the uplink and the downlink. 
	
	The achievable data rate of the communication link, $R_{\rm b}$, can be evaluated using the lower-bounded channel capacity formula for intensity modulation and direct detection (IM/DD) systems, assuming additive white Gaussian noise (AWGN) channel \cite{Lapidoth2009-sh}:
	\begin{equation}
		R_{\rm b}=\frac{1}{2}\log_2\left\{1+\frac{(\rho_{\rm pd}P_{\rm pd})^2}{2\pi e \sigma_{\rm n}^2}\right\},
	\end{equation}
	where $\rho_{\rm pd}$ (A/W) is the responsivity of the photodetector, $P_{\rm pd}$ is the received optical power at the photodetector (either $P_{\rm pd1}$ or $P_{\rm pd2}$), and $\sigma_{\rm n}^2$ is the noise variance at the receiver. The factor of 1/2 accounts for the on-off keying (OOK) modulation format. In practice, $R_{\rm b}$ may be lower due to effects such as modulation and demodulation, and channel impairments.
	
	The noise is composed of shot noise and thermal noise components. Shot noise arises from the statistical fluctuations of photon arrival and electron generation in the photodetector, while thermal noise is due to the random motion of electrons in the receiver circuit components. Note that the shot noise is related to the total light power, $P_{\rm pd}$, received by the PD, comprising the signal beam ($P_{\rm up}$ or $P_{\rm down}$) and the resudial beam ($P_{\rm res}=T_{\rm pd1}T'_{\rm M4}T_{\rm L5}T'_{\rm M1}P^-_{\rm 2\nu}$, only exists in uplink) in SHG process. The noise variance can be formulated as~\cite{Lapidoth2009-sh}:
	\begin{equation}
		\sigma_{\rm n}^2=2q_{\rm e}(\rho_{\rm pd}P_{\rm pd}+I_{\rm bk})B_{\rm r}+\frac{4kTB_{\rm r}}{R_{\rm IL}},
	\end{equation}
	where $q_{\rm e}$ is the elementary charge, $I_{\rm bk}=5100~{\mu\rm{A}}$ is the background-light-induced current, $B_{\rm r}$ is the bandwidth of the receiver, $k$ is Boltzmann constant, $T_{\rm t}$ is temperature, and $R_{\rm IL}$ is the load resistance of the receiver circuit.

	\section{Numerical Results}
	\label{sec:result}
	This section presents the numerical results obtained from our simulations based on the system model described in Section \ref{sec:model}. We will detail the parameters used in the simulations, the figures generated to analyze the system performance under various conditions, and a discussion of the key findings derived from these numerical experiments.
	
	\begin{table}[h]
		\centering
		\caption{Parameter Seting}
		\label{tab:params}
		\begin{tabular}{l l l}
			\hline
			Parameter & Symbol & Value \\
			\hline
			HR coating reflectivity & $R_{\rm hr}$ & 0.997 \\
			AR coating transmissivity & $T_{\rm ar}$ & 0.995 \\
			Distance  & $d_1$ & $(50-l_{\rm s}/n_{\rm s})$ \textrm{ mm} \\
			Distance  & $d_2$, $d_3$, $d_4$, $d_6$ & 50 \textrm{ mm} \\
			Distance  & $d_5$ & 97 \textrm{ mm} \\
			Focal length lens& $f_{\rm L1}$, $f_{\rm L2}$, $f_{\rm L4}$ & 50 \textrm{ mm} \\
			Focal length lens& $f_{\rm L3}$ & 100 \textrm{ mm} \\
			Length SHG crystal & $l_{\rm s}$ & 1 \textrm{ mm} \\
			Length gain medium & $l_{\rm g}$ & 1 \textrm{ mm} \\
			Length mirror M2 & $l_{\rm m}$ & 6 \textrm{ mm} \\
			Refractive index SHG crystal & $n_{\rm s}$ & 2.23 \\
			Refractive index gain medium & $n_{\rm g}$ & 1.96 \\
			Refractive index mirror M2 & $n_{\rm m}$ & 1.5 \\
			Refractive index concentrator & $n_{\rm c}$ & 1.5 \\
			Gain medium aperture radius & $a_{\rm g}$ & 1.4 \textrm{ mm} \\
			Fundamental wavelength & $\lambda_{\nu}$ & 1064 \textrm{ nm} \\
			Nonlinear coeff. SHG crystal & $d_{\rm eff}$ & 4.7 \textrm{ pm/V} \\
			Saturation intensity & $I_{\rm s}$ & $1.1976\times 10^7$~\textrm{W/m$^2$} \\
			Pump power & $P_{\rm in}$ & 60 \textrm{ W} \\
			Combined efficiency & $\eta_{\rm c}$ & 0.439 \\
			Responsivity PV panel & $\rho_{\rm pv}$ & 0.6 \textrm{ A/W} \\
			Diode saturation current & $I_0$ & 0.32 $\mu$\textrm{A} \\
			Number series PV cells & $N_{\rm s}$ & 1 \\
			Diode ideality factor & $n_{\rm d}$ & 1.48 \\
			Series resist. PV panel & $R_{\rm s}$ & 37~m$\Omega$ \\
			Shunt resist. PV panel & $R_{\rm sh}$ & 53.82~$\Omega$ \\
			Responsivity PD  & $\rho_{\rm pd}$ & 0.4 \textrm{ A/W} \\
			Receiver bandwidth & $B_{\rm r}$ & 800 \textrm{ MHz} \\
			Load resist. receiver & $R_{\rm IL}$ & 10 \textrm{ k$\Omega$} \\
			Background light current & $I_{\rm bk}$ & 5100 \textrm{$\mu$A} \\
			Vacuum permittivity & $\varepsilon_0$ & $8.854\times10^{-12}$ \textrm{ F/m} \\
			Speed of light & $c$ & $3\times10^8$ \textrm{ m/s} \\
			Boltzmann constant & $k$ & $1.38\times 10^{-23}$ \textrm{ J/K} \\
			Temperature & $T_{\rm t}$ & 300 \textrm{ K} \\
			Elementary charge & $q_{\rm e}$ & $1.602\times 10^{-19}$ \textrm{ C} \\
			Air attenuation coeff. & $\alpha_{\rm air}$ & $10^{-4}$ \textrm{m$^{-1}$} \\
			\hline
		\end{tabular}
	\end{table}
	
	\subsection{Parameters Settings}
	
	Here, lenses and crystals with AR coatings have a transmissivity of $T_{\rm ar}^2$ for both the fundamental (resonant) beam and the second harmonic beam, which include $T_{\rm L1}$~--~$T_{\rm L6}$, $T_{\rm g}$ and $T_{\rm s}$. As each surface of the gain medium has an AR coating, we have $T_{\rm gl}=T_{\rm gr}=T_{\rm ar}$. EOMs are HR coated for the fundamental beam, so we have $R_{\rm e1}=R_{\rm e2}=R_{\rm hr}$. However, for the second harmonic beam, it should pass through the front AR coating and then be reflected by the modulating surface, thus we have $R'_{\rm e1}=R'_{\rm e2}=T_{\rm ar}^2 R_{\rm hr}$. For frequency-selective mirrors, we have $T'_{\rm M4}=T'_{\rm M5}=T_{\rm ar}^2$ and $R_{\rm M5}=R_{\rm ar}$. The reflectivities of M2 and M3 at the fundamental frequency are investigated in subsection~\ref{sec:subRR}. M1 and M2 are anti-reflective for second harmonic beam, i.e., $T'_{\rm M1}=T'_{\rm M2}=T_{\rm ar}^2$. To provide the uplink carrier, M3 is partially reflective at the second harmonic frequency; thus, its reflectivity $R'_{\rm M3}$ is investigated in subsection~\ref{subsec:comm}. Unless otherwise specified, all remaining parameters are listed by default in Table~\ref{tab:params}~\cite{Moreira1997-rv, Quintana2017-ut, Long2020-jm}.

	\subsection{Stability Analysis}
	
	\begin{figure}[t]
		\centering
		\includegraphics[width=3.2in]{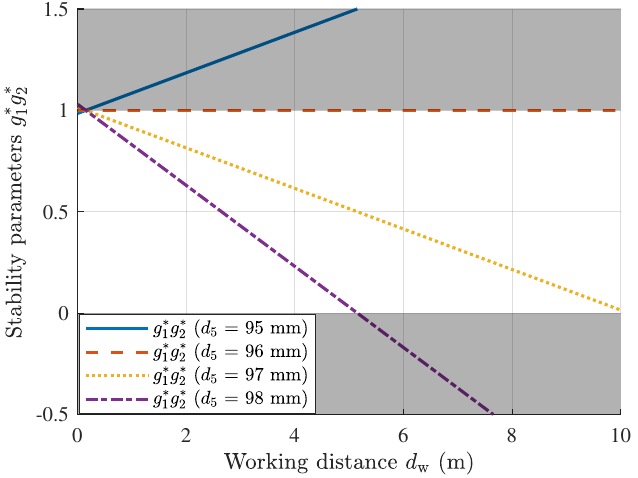}
		\caption{Stability as a function of working distance $d_{\rm w}$.}
		\label{fig:stability}
	\end{figure}
	
	The stability of the CSDR cavity is a critical factor in determining the system operational range and robustness. As depicted in Fig.~\ref{fig:stability}, the region of stable operation is fundamentally constrained within the range of $g_1^*g_2^* \in (0, 1)$. To investigate the impact of system geometry on stability, we analyze the relationship between the stability parameter $g_1^*g_2^*$ and the working distance $d_{\rm w}$ for various values of $d_5$.  As illustrated in Fig.~\ref{fig:stability}, the lines representing $g_1^*g_2^*$ as a function of $d_{\rm w}$ exhibit distinct slopes for each $d_5$ value.  Notably, these lines converge at a common intersection point around $d_{\rm w} = 148$~mm. This intersection suggests a critical working distance, below which stable operation is not consistently achievable across different $d_5\in\{96, 97, 98\}~\mbox{mm}$. Consequently, for the given system configuration, the working distance should ideally exceed $148$~mm to ensure stable resonator operation.
	
	Furthermore, Fig.~\ref{fig:stability} reveals a trend where increasing the working distance $d_{\rm w}$ leads to variable resonator stability. Namely, as $d_{\rm w}$ increases, the variance of the stability parameter product $g_1^*g_2^*$ progressively diminishes from $1$ towards $0$.   Theoretically, the maximum achievable working distance corresponds to the point where $g_1^*g_2^* = 0$. In the subsequent sections, we aim to design a $6$-m SLIPT system. Based on the stability analysis presented in Fig.~\ref{fig:stability}, we  set $d_5 = 97~\mbox{mm}$ to ensure the stable range covering the desired working distance.
	
	\subsection{Beam Radius Evolution}
	\label{subsec:radius}
	\begin{figure}[t]
		\centering
		\includegraphics[width=3.2in]{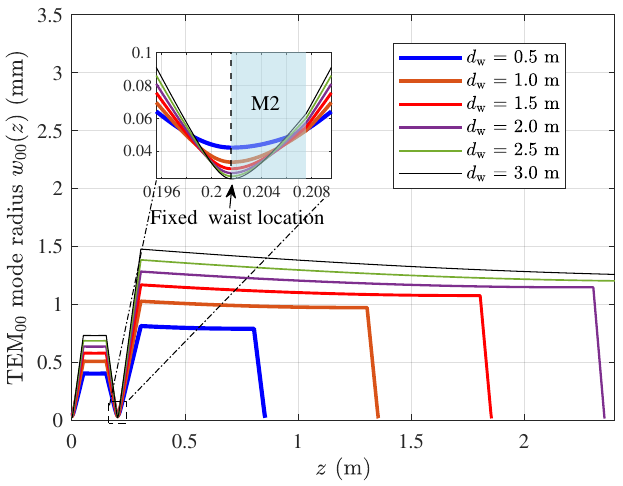}
		\caption{Evolution of the TEM$_{00}$ mode radius $w_{00}$ along the $z$-axis with varying working distances $d_{\rm w}$.}
		\label{fig:radius}
	\end{figure}
	
	Figure~\ref{fig:radius} presents a detailed analysis of the fundamental TEM$_{00}$ mode radius $w_{00}$ evolution along the resonator's z-axis for different working distances $d_{\rm w}$.  The turning points in the mode radius profile indicate the positions of the lenses L1 -- L4. Notably, $w_{00}$ within the intra-sub-resonator is approximately half that of the extra-sub-resonator. This size reduction is attributed to the lens pair (L2 and L3) acting as a beam-compressing telescope. As  $d_{\rm w}$ increases, a corresponding increase in $w_{00}$ throughout the cavity is observed.
	
	It is crucial to distinguish between the TEM$_{00}$ mode radius and the overall resonant beam radius.  As formulated in~\eqref{equ:M2}, if the TEM$_{00}$ mode is significantly smaller than the gain medium aperture, higher-order transverse modes can be excited. In such cases, the resonant beam radius at the gain medium approximates the aperture size. When many higher-order modes are present, diffraction losses decrease significantly because beam energy concentrates in lower-order modes. Conversely, if the TEM$_{00}$ mode margin approaches the edge of the gain medium aperture, the diffraction loss increases dramatically, leading to disruption of the resonance.
	
	Figure~\ref{fig:radius} confirms that the TEM$_{00}$ mode waist between lenses L2 and L3 is positioned at mirror M2 consistently, even with varying $d_{\rm w}$, which is critical as shifts in the waist location could lead to mode mismatch between the intra- and extra-sub-resonators, resulting in substantial diffraction loss and limiting the effective working range. This fixed waist location at M2 is a result of deliberate design choices in this deployment. For alternative system parameters (e.g., changes in $d_1$ or $l_{\rm s}$), it is imperative to first identify the waist location between L2 and L3 and then position mirror M3 at this waist to minimize mode mismatch losses across the working range.
	
	\subsection{Beam Power and Working Distance}

	\begin{figure}[t]
		\centering
		\includegraphics[width=3.2in]{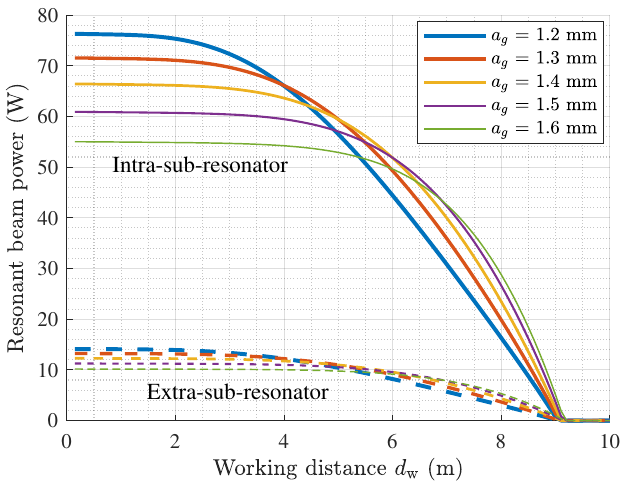}
		\caption{Resonant beam power $P_{\nu}$ within the intra-sub-resonator and $P_{\rm \nu,ext}$ within the extra-sub-resonator as a function of working distance}
		\label{fig:powernu}
	\end{figure}
	\begin{figure}[t]
		\centering
		\includegraphics[width=3.2in]{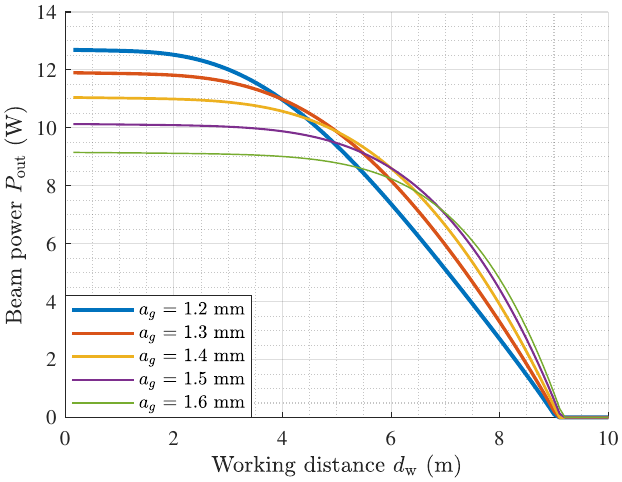}
		\caption{Resonant beam power $P_{\rm out}$ output by M3 as a function of working distance}
		\label{fig:output}
	\end{figure}

	Figures~\ref{fig:powernu} show that the generated resonant beam power, both in intra-sub-resonator $P_{\nu}$ and in extra-sub-resonator $P_{\rm \nu,ext}$, clearly depends on the working distance $d_{\rm w}$, thus validating the analytical predictions in subsection~\ref{subsec:radius}. At short working distances, the beam power remains relatively stable. This is because at small $d_{\rm w}$, the  fundamental TEM$_{00}$ mode radius is significantly smaller than the gain medium aperture, rendering diffraction loss negligible. However, as $d_{\rm w}$ continues to increase, exceeding a critical threshold, the resonant beam power as well as the output power reduction intensifies rapidly, ultimately diminishing to zero.
	
	The resonant beam power in the intra-sub-resonator is substantially higher than that in the extra-sub-resonator, exhibiting an approximate 3:1 ratio. This highlights the advantage of CSDRs in reducing the free-space power exposure, which guarantees an improved  safety and transmision efficiency. For instance, with a 0.1 reflectivity mirror M3, a $10\mbox{-W}$ resonant beam output implies an approximately $11.1\mbox{-W}$ forward-traveling power and $1.1\mbox{-W}$ backward-reflected power in the extra-sub-resonator (free space section). This power level is notably lower than that within the intra-sub-resonator, i.e., about $\mbox{60~W}$. High intra-sub-resonator power enhances the SHG efficiency, and low extra-sub-resonator power guarantees safer operation, leading to increased permissible transmission power within safety limits. Because this coupled design reduces the intra-cavity power present in the free space, it provides higher transmission efficiency than a conventional non-coupled SDR, especially when the air condition is bad.

	Besides, Fig.~\ref{fig:powernu} and Fig.~\ref{fig:output} reveals that reducing the gain medium radius $a_{\rm g}$ (the pump beam radius is also reduced) leads to a relative increase in beam power at shorter $d_{\rm w}$. Conversely, the onset of rapid beam power decrease shifts to shorter distances, and the beam power at longer $d_{\rm w}$ becomes lower compared to configurations with a larger $a_{\rm g}$. This behavior suggests that an appropriate $a_{\rm g}$ is crucial to ensure a desired transmission power level at a specific working distance. Therefore, optimizing $a_{\rm g}$ is a trade-off relying on the intended application and operational range.

	\subsection{Charging Power \textit{vs.} $R_{\rm M2}$ and $R_{\rm M3}$}
	\label{sec:subRR}
	
	\begin{figure}[t]
		\centering
		\includegraphics[width=3.3in]{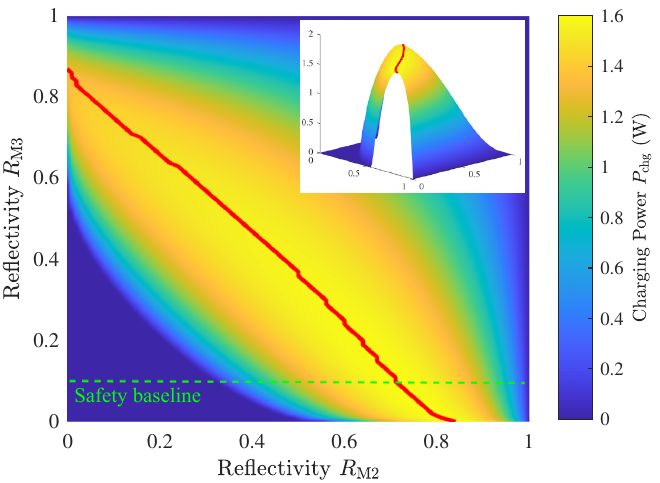}
		\caption{Charging power variation with the reflectivity of M2 and M3}
		\label{fig:charging}
	\end{figure}
	
	As depicted in Fig.~\ref{fig:charging}, the charging power provided by the PV panel is influenced by varying configurations of reflectivities $R_{\rm M2}$ and $R_{\rm M3}$.  The figure reveals a spindle-shaped region characterized by high power. The red line in Fig.~\ref{fig:charging} delineates the trend of maximum achievable power, connecting the peak power values obtained for each distinct $R_{\rm M3}$ configuration. This trend shows an inverse relationship: lower $R_{\rm M3}$ values correspond to maximum power at higher $R_{\rm M2}$, and vice versa. As $R_{\rm M3}$ increases beyond a certain threshold, the corresponding $R_{\rm M2}$ on the red line gradually decreases to zero. The power values along this line are comparable, suggesting that a wide range of \{$R_{\rm M2}$, $R_{\rm M3}$\} pairs for choosen. This implies a strategy of selecting a smaller $R_{\rm M3}$ with a larger $R_{\rm M2}$ to ensure a reduced beam power in free space while maintaining the output charging power, thereby enhancing safety. Therefore, for a desired charging power level, establishing a safety baseline for $R_{\rm M3}$ is advisable.  We at first set $R_{\rm M3}$ below this baseline, and then optimize  $R_{\rm M2}$ to attain the maximum power. For example, with a safety baseline of $R_{\rm M3}=0.1$, an $R_{\rm M2}$ of $0.72$ can be used. Notably, this safety baseline is chosen for demonstrative purposes only. In practice, the safety transmission power should be estimated by light exposure measurement and analysis, which exhibit variations in different systems.
	
	\subsection{Uplink and Downlink Rates \textit{vs.} $R'_{\rm M3}$}
	\label{subsec:comm}
	
	\begin{figure}[t]
		\centering
		\includegraphics[width=3.2in]{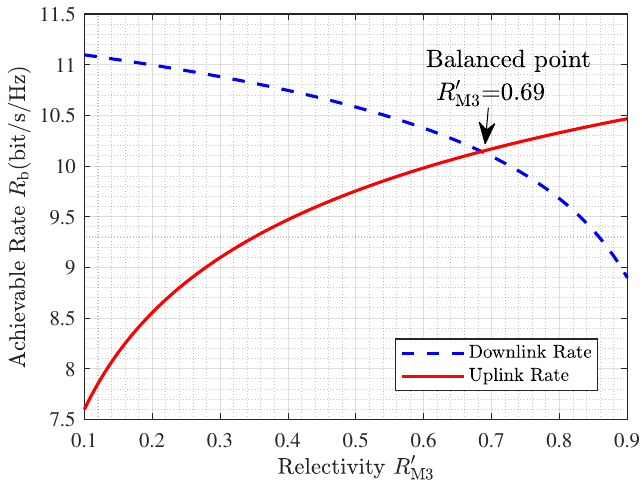}
		\caption{Achievable rate of uplink and downlink communications variation with the reflectivity of M3 at the second-harmonic frequency}
		\label{fig:rate}
	\end{figure}
	
	Figure~\ref{fig:rate} shows the achievable rates (uplink and downlink) as functions of the reflectivity $R'_{\rm M3}$ at the second-harmonic frequency. The downlink rate exhibits a decreasing trend with increasing $R'_{\rm M3}$, starting from approximately $11.2$ bit/s/Hz at $R'_{\rm M3} = 0.1$ and declining to around $9$ bit/s/Hz at $R'_{\rm M3} = 0.9$. In contrast, the uplink rate demonstrates an inverse relationship, increasing from $7.6$ bit/s/Hz to $10.5$ bit/s/Hz as $R'_{\rm M3}$ grows. An intersection point is observed at approximately $R'_{\rm M3} = 0.69$, indicating balanced communication performance for uplink and downlink at this reflectivity. It can be seen that the balanced $R'_{\rm M3}$ is greater than 0.5, reflecting the non-symmetry induced by shot noise originating from the leftward-propagating residual beam. This observation highlights the role of $R'_{\rm M3}$ in power allocation and consequently influencing the achievable data rates in both directions.

	\section{Conclusions}
	\label{sec:con}
	
	This paper proposed a novel simultaneous light information and power transfer~(SLIPT) system based on a coupled spatially distributed resonator~(CSDR) structure and second harmonic generation scheme, which reduces the power exposure in the free space, enhance the power transmission efficiency and safety, and provides bidirectional communications. 
	The numerical experiment result demonstrate the stable region, intra-cavity power, output power, charging power, and achievable communication rate. The findings not only demonstrate the feasibility of stable operation and efficient power transfer across varying distances, but also highlights trade-offs in reflectivity selection within a CSDR. This research offers significant insights for the design and optimization of CSDR-based resonant beam systems, advancing the field of SLIPT.
	

	
	%

	%

	\section*{Acknowledgment}
	The authors would like to acknowledge Gemini for grammar checking and text polishing and Grok for generating the background image in Fig.~\ref{fig:demo}. The generated contents were carefully checked and confirmed by the authors.


	\ifCLASSOPTIONcaptionsoff
	\newpage
	\fi

	
	
	
	\bibliographystyle{IEEETran}
	\small
	%
	\bibliography{mybib}

	%
	%
	
	%
	
	%
	%
	
	
	
	
	

	
\end{document}